\documentclass[conference]{IEEEtran}
\usepackage{amsmath,amssymb,eucal,graphicx}
\usepackage{pst-plot}
\usepackage{epsfig}
\usepackage{amsthm}
\usepackage{amsfonts}
\usepackage{epsfig}
\usepackage{float}
\usepackage{pstricks}
\usepackage{color}
\usepackage{multicol,subfigure}
\usepackage{mathcomSTEP,stmaryrd}

\theoremstyle{remark}

\begin{document}

\title{
A New Random Coding Technique that Generalizes Superposition Coding and Binning
}
\author{%
\IEEEauthorblockN{Stefano Rini}
\IEEEauthorblockA{
Lehrstuhl f\"{u}r Nachrichtentechnik \\
Technische Universit\"{a}t M\"{u}chen \\
Arcisstra{\ss}e 21, 80333 M\"{u}nchen, Germany
%
\\ Email: \tt stefano.rini@tum.de}}

\maketitle

\begin{abstract}

Proving capacity for networks without feedback or cooperation usually involves two fundamental random coding techniques: superposition coding and binning.
%
%
Although conceptually very different, these two techniques often achieve the same performance, suggesting an underlying similarity.
%
%
In this correspondence we propose a new random coding technique that generalizes superposition coding and binning
and provides new insight on relationship among the two
%
%
%
%
With this new theoretical tool, we derive new achievable regions for three classical information theoretical models: multi-access channel, broadcast channel, the interference channel, and show that, unfortunately, it does not improve over the largest known achievable regions for these cases.
\end{abstract}

{\IEEEkeywords
random coding, superposition coding, binning, multi access channel, broadcast channel, interference channel.
}

\section{Introduction}

\begin{figure}
\centering
\includegraphics[width=8.5 cm]{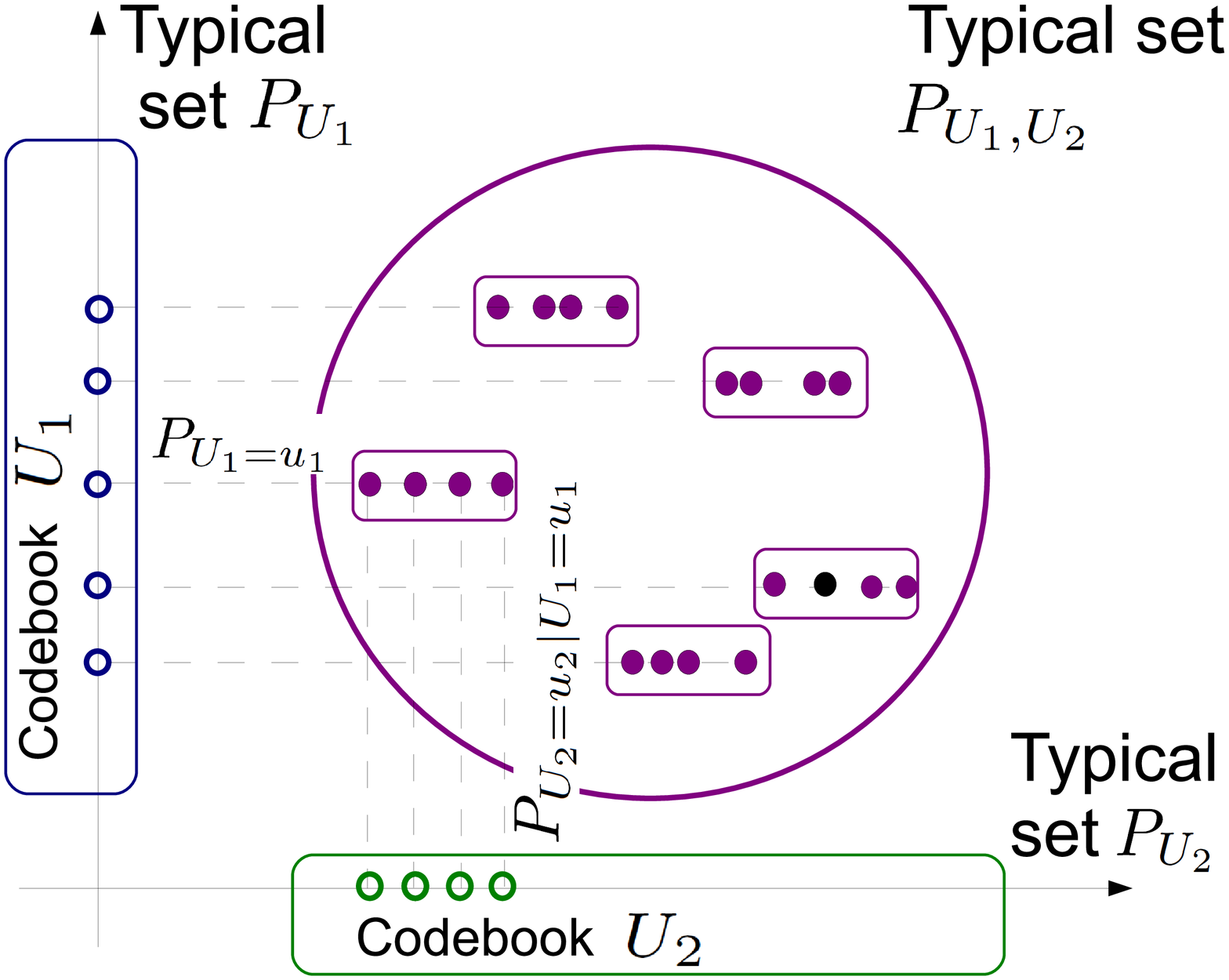}
\vspace{-.5 cm}
\caption{A graphical representation of superposition coding.}
\label{fig:SPCcodeword}
\vspace{-.8 cm}
\end{figure}

Apart from few notable exceptions \cite{sridharan2008capacity,abbe2010mac}, the capacity region for a general multi-terminal network
is shown using random coding techniques such as rate splitting, time sharing, superposition coding, binning, Markov encoding, quantize, and forward and few others.
For networks with no feedback or cooperation, the two random coding techniques are usually considered when proving capacity: superposition coding and binning.
%
%
\emph{Superposition coding} can be intuitively be thought of as stacking codewords on top of each other
\cite{cover1972broadcast} and is obtained by generating the codewords of the ``top'' codebook conditionally dependent on the ``base'' codeword.
%
%
A typical representation of this encoding technique is the one Fig. \ref{fig:SPCcodeword} \cite[Fig 4]{cover1972broadcast}
where the codeword $U_{2}^N$ is superposed to the codewords $U_1^N$.
A codeword $U_{1}^N$ in the base codebook is randomly selected from the typical set $\Tcal_{\ep}^N (P_{U_1})$.
For each base codeword, a top codebook is generated by selecting random elements from the typical set $\Tcal_{\ep}^N \lb P_{U_2 |U_1}\rb $.
%
%
%
%
%
%
Superposition coding is often thought of as placing spheres, or clouds, in the typical set $\Tcal_{\ep}^N(P_{U_1 U_2})$ : the base codewords are ``cloud centers'' while top codewords are ``satellite codewords''.
If the number of spheres is small enough--a low rate for the base codewords--and their size is small enough--a low rate for the top codewords--then codewords are sufficiently spaced apart to allow successful decoding.
%
%

\begin{figure}
\centering
\vspace{-.6 cm}
\includegraphics[width=8.5 cm]{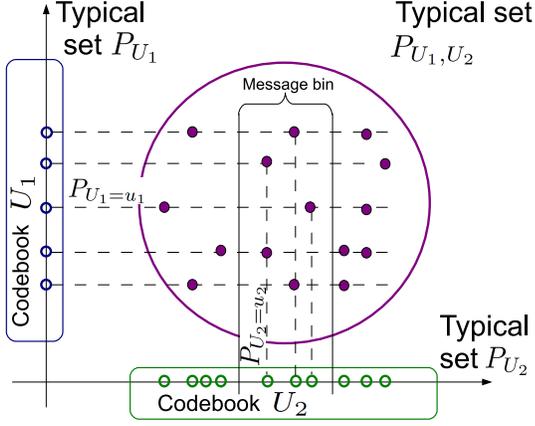}
\vspace{-.5 cm}
\caption{A graphical representation of joint binning.}
\label{fig:DPCcodeword}
\vspace{-.8 cm}
\end{figure}

\emph{Binning} \footnote{sometimes referred to as Cover's random binning \cite{ElGamalLectureNotes} or Gel'fand-Pinsker coding~\cite{GelFandPinskerClassic}.}
 allows a transmitter to ``pre-cancel'' (portions of)  the interference experienced at a receiver.
%
%
A usual representation of binning \cite[Fig. 14.19]{cover1991elements} is the one in Fig. \ref{fig:DPCcodeword}: here
the codewords $U_2^N$ is binned against $U_1^N$. The codebook of $U_1^N$ is generated as in superposition coding while
the codewords for $U_2^N$ and are selected from the typical set $\Tcal_{\ep}^N(P_2)$ and placed in bins.
Codewords in the same bin are associated with the same message, i.e. multiple codewords can be used to communicate the same message.
The codeword  $U_2^N$ in the bins are selected for transmission when it belongs to the typical set $\Tcal_{\ep}^N \lb P_{U_1, U_2} \rb$,
even though generated independently from $U_1^N$.
It is possible to find a codeword that satisfies this condition if the size of each bin is sufficiently large.
%
%
%
%
%
%
%
%
%
%
Binning is commonly interpreted as dividing the typical set $\Tcal_{\ep}^N \lb P_{U_1, U_2} \rb$ in the partitions
formed by the bins in which the codewords $U_1^N$ and $U_2^N$  are placed.
Encoding is successful when the size of the bins is sufficiently large--large binning rate--while decoding is successful if the transmitted codewords are sufficiently far apart--low message rate.
In certain cases, it is possible to simultaneously bin two codewords against each other: this coding technique is usually referred to as \emph{joint binning}.
%
%
%
%
%
%
%

%
%
%
Despite the difference in this two random coding techniques, in many cases they have identical performance \cite[Sec. VI]{rini2011achievable}:
this suggests an underlying similarity in the way the two techniques select the codewords to be transmitted.
To gain a better understanding of the properties of theses strategies, we develop a new random coding technique that encompasses superposition coding and binning as special cases.
In this scheme codewords are first superposed according to a certain distribution, the \emph{codebook distribution}, and successively binned to appear as if generated according to a different distribution, the \emph{encoding distribution}.
%
%
%
Classical superposition coding corresponds to the case where the binning distribution is the same as the codebook distribution
while binning is obtained when the codebook distribution has independent codewords.
All the strategy in between these two cases have never been previously considered in literature.
%
We use this new random coding technique to derive achievable regions for the multi-access channel, the broadcast channel, and the interference channel.
Unfortunately these new achievable regions do not improve on the largest known achievable regions for the broadcast channel and the interference channel but show that these regions can be obtained with a wider set of encoding strategies than what was previously.

{\bf Paper Organization:}
Section~\ref{sec:Combining Superposition Coding and Binning} introduces the a new random coding techniques that generalizes superposition coding and binning.
Section~\ref{sec:Achievable Regions for Classical Channel Models} presents new achievable regions for classical communication models.
Section~\ref{sec:conclusion} concludes the paper.

\section{Combining Superposition Coding and Binning}
\label{sec:Combining Superposition Coding and Binning}

We introduce the new random coding technique that generalizes superposition coding and binning with a simple example.
Consider a classical Broadcast Channel with a common message (BC-CM) where two messages $W_1$ and $W_2$ are encoded at transmitter~1, message $W_2$ is decoded at receiver~2 while message $W_1$ is decoded  both decoder~1 and decoder~2.
The channel outputs $(Y_1,Y_2)$ are obtained from the channel input $X$ from the channel transition probability $P_{Y_1,Y_2 | X}$.

Take any distribution for the codebook generation and the encoding procedure
\ea{
P_{\rm codebook}
=P_{U_1^{\rm c} U_2^{\rm c}}, \quad \quad
P_{\rm encoding}
=P_{U_1^{\rm e} U_2^{\rm e}},
\label{eq:example encoding codebook distribution}
}
and let $w_i \in \lsb 1 \ldots 2^{N R_i} \rsb $ be the messages $W_i$  to be transmitted for $i \in \{1,2\}$.
%

\noindent
$\bullet$ {\bf Codebook Generation}

\noindent
1) \
Generate $2^{L_1}$ codewords $U_1^N$ with $N$ iid draws from the distribution $P_{U_1^{\rm c}}$.
Index these codewords as $U_{1}^N \lb w_1, b_1 \rb$
for  $b_1\in  \lsb 1 \ldots2^{\Ro_1}\rsb$
with  $L_1 =R_1 +\Ro_1$ .

\noindent
2) \
For each $ U_1^N \lb w_1, b_1 \rb$ , generate $2^{N L_2}$  codewords $U_{2}^N$ with $N$ iid draws from the distribution
$P_{U_{2}^{\rm c} | U_1^{\rm c}}$.
Index these codewords as $U_{2}^N \lb w_{2}, b_{2},  w_{1}, b_{1} \rb$
for  $b_{2} \in \lsb 1 \ldots2^{\Ro_{2}} \rsb $ with  $L_{2}=R_{2}+\Ro_{2}$

\noindent
$\bullet$ {\bf Encoding Procedure}

\noindent
For each message set $(w_1,w_2)$,  choose the bin indexes $b_1$ and $b_2$ so that
\ea{
\lcb U_{1}^N (w_1, b_1),  U_2^N (w_2, b_2 , w_1, b_1)\rcb  \in \Tcal_\ep^N ( P_{U_1^{\rm e} U_2^{\rm e}}).
\label{eq:example encoding condition}
}
If no such set $(b_1,b_2)$ exists, pick two indexes at random.
%
Generate the channel input $X$ as a deterministic function of the Random Variables (RVs) $U_1$ and $U_2$.

\noindent
$\bullet$ {\bf Decoding Procedure}


\noindent
1) \
Decoder~1 looks for a set of indexes $\wh_1^1, \bh_1^1$ such that
  $$
  \lcb Y_1^N , U_1^N (\wh_1^1 , \bh_1^1) \rcb \in \Tcal_{\ep}^N ( P_{Y_1,U_1^{\rm e}})
  $$
2) \
Decoder~2 looks for a set of indexes $\wh_1^2, \wh_2^2, \bh_1^2, \bh_2^2$
    $$
  \lcb Y_2^N , U_1^N (\wh_1^2 , \bh_1^2) , U_2^N (\wh_1^2 , \bh_1^2,\wh_2^2 , \bh_2^2)\rcb \in \Tcal_{\ep}^N ( P_{Y_2 U_1^{\rm e} U_2^{\rm e}})
  $$


To determine the performance of the achievable scheme above we need to determine the encoding and decoding error probabilities.

The key techniques in bounding the encoding error is related to the probability that there exists a random vector
 $U_1^N$ in the typical set of certain distribution  $\Tcal_{\ep}^N(P_{U_2})$, that also belongs  to the typical set of a different distribution $\Tcal_{\ep}^N ( P_{U_2} )$.
The probability of this event can be bounded using \cite[Lem. 2.6 ]{CsiszarKornerBook}.
\begin{lem} {\bf Inaccuracy, \cite[Lem. 2.6]{CsiszarKornerBook}.}
\label{lem:Inaccuracy}
Consider a two general distributions $P_{U_2}$ and $P_{U_2}$ and $u_1^N \in \Tcal_{\ep}^{N}(P_{U_1})$, then
$$
P_{U_2}^N(u_1^N) = \exp \lcb -N (D(P_{U_1}||P_{U_2})+H(U_1)\rcb
$$
\end{lem}
The quantity
$$
D(P_{U_1}||P_{U_2})+H(U_1)= - \sum_{u_1 } P_{U_1}(u_1) \log (P_{U_2} (u_1 ))
$$
is referred to as \textit{inaccuracy} and it can be used to derive a more general version of the covering lemma \cite{ElGamalLectureNotes}.

\begin{thm} {\bf Generalized Covering Lemma}
\label{th:Generalized Covering Lemma}

Consider the set of  RVs $U_1$ and $U_2$ with distribution $P_{U_1 U_2}$
and take $2^{N \Ro}$ i.i.d. sequences $u_2^N$ from the typical set $\Tcal_{\ep}^N(P_{U_2})$ , indexed as $u_2^N(i), \ i \in \{1 \ldots 2^{n \Ro}\}$, then there
exists $\delta(\ep) \goes 0$  as $\ep \sgoes 0$ such that
$$
P \lsb \exists \ \ih \ \in \{1 \ldots 2^{n \Ro}\}, \ \ST u_2^N( \ih ) \in \Tcal_{\ep}^N(P_{U_1})  \rsb \goes 1,
$$
as $N \goes \infty $ if
$$
\Ro  \geq D(P_{U_1} || P_{U_2} )+\delta(\ep).
$$
\end{thm}

\begin{IEEEproof}
 The complete proof is provided in \cite{rini2012Extension}.
\end{IEEEproof}

With Th. \ref{th:Generalized Covering Lemma} we derive the achievable region of the proposed random coding strategy.

\begin{cor}{\bf An Achievable Region for the BC-CM}
\label{th:A brief example}
The following region is a achievable for a general BC-CM
\ea{
\Ro_1 + \Ro_2 &  \geq  D( P_{U_1 , U_2}^e|| P_{U_1, U_2}^c ) \\
L_1 \quad \quad  \ & \leq I(Y_1; U_1^{\rm e}) + D(P_{U_1^{\rm e}} || P_{U_1^{\rm c}}) \nonumber \\
L_2  & \leq I(Y_2; U_2^e | U_1^{\rm e} ) +  D(P_{U_2^{\rm e}} || P_{U_2^{\rm c}}| P_{U_1^{\rm e}}) \nonumber  \\
L_1+L_2 & \leq I(Y_2; U_1^{\rm e} , U_2^{\rm e})+ D( P_{U_1^{\rm e} U_2^{\rm e}} || P_{U_1^{\rm c} U_2^{\rm c}}), \nonumber
}
for
$
L_i=R_i + \Ro_i 
$
for $i \in \{1,2\}$ and any distribution of $P_{\rm codebook}$ and $P_{\rm encoding}$ in \eqref{eq:example encoding codebook distribution}.
\end{cor}

\begin{IEEEproof}
 The complete proof is provided in \cite{rini2012Extension}.
\end{IEEEproof}


\noindent
 { $\bullet$ \bf Superposition coding:}
 is obtained by having the distribution imposed at encoding  $P_{\rm encoding}$ equal the distribution of the codebook
 $P_{\rm codebook}$, i.e.
 \ea{
 P_{U_1^{\rm e} U_2^{\rm e}}= P_{U_1^{\rm c} U_2^{\rm c}}.
 }
In this case the binning rates $\Ro_1$ and $\Ro_2$ can be set to zero, thus obtaining the region
 \ea{
 R_1\quad \quad  \ & \leq I(Y_1 ; U_1^{\rm c}) \nonumber \\
 R_2 & \leq I(Y_1 ; U_2^{\rm c} | U_1^{\rm c}) \nonumber \\
 R_1 + R_2 & \leq I(Y_1 ; U_1^{\rm c} , U_2^{\rm c}),
 }

 \noindent
 { $\bullet$  \bf Joint Binning:}
corresponds to the case where the distribution of the codebook $P_{\rm codebook}$ equals the product of the marginals of
encoding distribution:
   \ea{
  P_{U_1^{\rm c} U_2^{\rm c}}= P_{U_1^{\rm e}} P_{U_2^{\rm e}}.
  }
which results in the region
  \ea{
  \Ro_1+\Ro_2 & \geq I(U_1^{\rm e}; U_2^{\rm e}) \nonumber \\
  L_1  \quad  \quad \         & \leq I(Y_1 ; U_1^{\rm e}) \nonumber \\
  L_2         & \leq I(Y_2; U_2^{\rm e} | U_1^{\rm e}) \nonumber \\
  L_1+L_2         & \leq I(Y_2; U_1^{\rm e} , U_2^{\rm e}) +I(U_1^{\rm e}; U_2^{\rm e}),
  }

\noindent
{ $\bullet$  \bf Binning:} The region with binning  is obtained from the region with joint binning of  above
by setting either $\Ro_1$ or $\Ro_2$ to zero.


\section{Achievable Regions for Classical Channel}
\label{sec:Achievable Regions for Classical Channel Models}

We apply the new random coding technique in Sec. \ref{sec:Combining Superposition Coding and Binning}
to the Multi-Access Channel with Common Messages (MAC-CM) \cite{ahlswede1971multi,liao1972multiple},
 the Broadcast Channel (BC) \cite{cover1972broadcast,bergmans1973random,gallager1974capacity}
 and the InterFerence  Channel (IFC) \cite{ahlswede1974capacity,Han_Kobayashi81}.
Capacity is known for the MAC-CM and for a subsets of both the BC and the IFC.
The largest achievable regions in each case can be achieved by employing a combination of rate splitting,superposition coding and binning \cite[Sec. VI]{rini2011achievable}.
In the following we adopt the notation in \cite{rini2011achievable} to describe the channel model and distribution of messages and codewords.
In particular, the codeword $U_{\iv \sgoes \jv}^N$, with rate $R_{\iv \sgoes \jv}$, encodes the messages $W_{\iv \sgoes \jv}$ from the
$\iv$ set of transmitters to the $\jv$ set of receivers.

\subsection{The Multi-Access Channel with Common Messages}
\label{sec:MAC}
In the classical MAC \cite{ahlswede1971multi,liao1972multiple}, two transmitters communicate a message each to a single decoder.
In the MAC-CM an additional common message is transmitted by each source to the decoder \cite{slepian1973coding}.
%
Let $U_{\iv \sgoes 1}^N$ be the codeword associated with the message from transmitter $\iv$ to receiver~1, for $\iv \in \lcb 1,2, \{1,2\} \rcb $ respectively.
%
Using the random coding technique in Sec. \ref{sec:Combining Superposition Coding and Binning},
we can superpose $U_{1 \sgoes 1}^N$ and $U_{2 \sgoes 2}^N$ over $U_{ \{1,2\} \sgoes 1}^N$
and successively bin $U_{1 \sgoes 1}^N$ and $U_{2 \sgoes 2}^N$ against  $U_{ \{1,2\} \sgoes 1}^N$ .
%
\begin{cor}{\bf An Achievable Region for the MAC-CM}
\label{cor:An Achievable Region for the MAC}
The following region is achievable for a general MAC-CM:
\ea{
\Ro_{ 1 \sgoes 1} & \geq D ( P_{U_{1 \sgoes 1}^{\rm e}} || P_{U_{1 \sgoes 1}^{\rm c}} | P_{U_{1 \sgoes \{1,2\}}^{\rm c}})  \nonumber \\
\Ro_{ 2 \sgoes 1} & \geq D ( P_{U_{2 \sgoes 1}^{\rm e}} || P_{U_{2 \sgoes 1}^{\rm c}} | P_{U_{1 \sgoes \{1,2\}}^{\rm c}})  \nonumber \\
 R_{\{1,2\} \sgoes 1} + L_{1 \sgoes 1} + L_{2 \sgoes 1} & \leq I(Y_1 ;  U_{1 \sgoes {1,2}}^{\rm e}, U_{1 \sgoes 1}^{\rm e}, U_{2 \sgoes 1}^{\rm e})  \nonumber \\
 & \quad \quad + D \lb P_{\rm encoding} | P_{\rm codebook} \rb \nonumber \\
 L_{1 \sgoes 1} + L_{2 \sgoes 1}  &  \leq  I(Y_1 ; U_{1 \sgoes 1}^{\rm e}, U_{2 \sgoes 1}^{\rm e} | P_{U_{ \{1,2\} \sgoes 1}^{\rm e}}) \nonumber \\
& \quad \quad + D \lb P_{\rm encoding} | P_{\rm codebook} \rb \nonumber \\
 L_{1 \sgoes 1} \quad \quad  \quad \ & \leq  I(Y_1 ; U_{1 \sgoes 1}^{\rm e} | U_{ \{1,2\} \sgoes 1}^{\rm e}, U_{ 2 \sgoes 1}^{\rm e}) \nonumber \\
 & \quad \quad + D ( P_{U_{1 \sgoes 1}^{\rm e}} || P_{U_{1 \sgoes 1}^{\rm c}} | P_{U_{1 \sgoes \{1,2\}}^{\rm c}})  \nonumber \\
 L_{2 \sgoes 1}  & \leq  I(Y_1 ; U_{2 \sgoes 1}^{\rm e} | U_{ \{1,2\} \sgoes 1}^{\rm e}, U_{ 1 \sgoes 1}^{\rm e}) \nonumber \\
  & \quad \quad + D ( P_{U_{1 \sgoes 2}^{\rm e}} || P_{U_{1 \sgoes 2}^{\rm c}} |P_{U_{1 \sgoes \{1,2\}}^{\rm c}}),
  \label{eq:mac achievable region}
}
union over all the distributions that  factor as
\ea{
P_{\rm codebook} &= P_{U_{ \{1,2\} \sgoes 1}^{\rm c}} P_{U_{1 \sgoes 1 }^{\rm c} | U_{ \{1,2\} \sgoes 1}^{\rm c}} P_{U_{2 \sgoes 1 }^{\rm c} | U_{ \{1,2\} \sgoes 1}^{\rm c}} \nonumber \\
P_{\rm encoding} &= P_{U_{ \{1,2\} \sgoes 1}^{\rm c}} P_{U_{1 \sgoes 1 }^{\rm e} | U_{ \{1,2\} \sgoes 1}^{\rm c}} P_{U_{2 \sgoes 1 }^{\rm e} | U_{ \{1,2\} \sgoes 1}^{\rm c}},
\label{eq:distribution MAC encoding}
}
and for $L_{\iv \sgoes \jv}=R_{\iv \sgoes \jv}+\Ro_{\iv \sgoes \jv}$.
\end{cor}

\begin{IEEEproof}
 The complete proof is provided in \cite{rini2012Extension}.
\end{IEEEproof}

After the Fourier-Motzkin Elimination (FME) of the region in \eqref{eq:mac achievable region}, we obtain the classical region  \cite{slepian1973coding}
 \ea{
 R_{\{1,2\} \sgoes 1 } + R_{1 \sgoes 1}+R_{2 \sgoes 1}  & \leq I(Y_1 ; U_{\{1,2\} \sgoes 1 }^{\rm c}, U_{1\sgoes 1 }^{\rm e}, U_{2\sgoes 1 }^{\rm e}) \nonumber \\
  R_{1 \sgoes 1}+R_{2 \sgoes 1}  & \leq I(Y_1 ; U_{1\sgoes 1 }^{\rm e}, U_{2\sgoes 1 }^{\rm e}| U_{\{1,2\} \sgoes 1 }^{\rm e} ) \nonumber \\
  R_{1 \sgoes 1} \quad \quad \quad \  & \leq I(Y_1 ; U_{1\sgoes 1 }^{\rm e} | U_{\{1,2\} \sgoes 1 }^{\rm c} , U_{2 \sgoes 1 }^{\rm e}) \nonumber \\
  R_{2 \sgoes 1} & \leq I(Y_1 ; U_{2\sgoes 1 }^{\rm e} | U_{\{1,2\} \sgoes 1 }^{\rm c} , U_{1 \sgoes 1 }^{\rm e}),
\label{eq:mac achievable region 1}}
union over all the possible distributions in \eqref{eq:distribution MAC encoding} which is indeed capacity.
Cor. \ref{cor:An Achievable Region for the MAC} shows that the capacity  of the MAC-CM can be achieved with any distribution of the codewords $U_{1\sgoes 1}^N$ and
$U_{2 \sgoes 1}^N$ for as long as the codewords can be further binned to impose the distribution of the matching outer bound.
%
The distance between the codewords at generation and after encoding has no effect on the resulting achievable scheme.

\subsection{The Broadcast Channel}
\label{sec:BC}
In the BC \cite{cover1972broadcast,bergmans1973random,gallager1974capacity} one encoder wants to communicate to two decoders a message each.
For this channel model, rate splitting can be applied so as to split each message in private and common part;
the two common part can then be embedded into a single common message.
%
%
This transforms the problem of achieving the rate vector $ [ R_{1 \sgoes 1}',R_{1 \sgoes 2}' ]$ in the problem of achieving the rate vector
$[ R_{1 \sgoes 1},R_{1 \sgoes 2},R_{1 \sgoes \{1,2\}} ]$ where
\ea{
\lsb \p {
R_{1 \sgoes 1}' \\
R_{1 \sgoes 2}'
}\rsb
=
\lsb \p{
1 & 0 & \al \\
0 & 1 & \alb \\
}\rsb
\lsb R_{1\sgoes 1 } \ R_{1\sgoes 2} \ R_{1\sgoes \{1,2\} }\rsb^T,
\label{eq:rate splitting BC}}
for any $\al \in [0 \ldots 1]$ and $\alb=1-\al$.
The random coding technique of Sec. \ref{sec:Combining Superposition Coding and Binning}
can be applied to the BC after the rate splitting in \eqref{eq:rate splitting BC} by superposing
$U_{1 \sgoes 1 }^N$ and $U_{1 \sgoes 2}^N$ over $U_{1 \sgoes \{1,2\}}^N$ and
successively jointly binning $U_{1 \sgoes 1}^N,U_{1 \sgoes 2}^N$ and $U_{1\sgoes \{1,2\}}^N$.
\begin{cor}{\bf An Achievable Scheme for the BC}
\label{cor:An Achievable Scheme for the BC}
The following region is achievable for a general BC:
\ea{
 \Ro_{1 \sgoes \{1,2\}}  +\Ro_{1\sgoes 1} & +\Ro_{1\sgoes 2}   \geq  \nonumber \\
&  D( P_{U_{1 \sgoes 1}^{\rm e} U_{1 \sgoes 2}^{\rm e} U_{1 \sgoes \{1,2\}}^{\rm e} }|| P_{U_{1 \sgoes 1}^{\rm c} U_{1 \sgoes 2}^{\rm c} U_{1 \sgoes \{1,2\}}^{\rm c} })& \nonumber \\
 \Ro_{1 \sgoes \{1,2\}} + \Ro_{1 \sgoes 1}     & \geq  D(P_{ U_{1 \sgoes 1}^{\rm e} U_{1 \sgoes \{1,2\}}^{\rm e} } ||P_{ U_{1 \sgoes 1}^{\rm c} U_{1 \sgoes \{1,2\}}^{\rm c} } )\nonumber \\
 \Ro_{1 \sgoes \{1,2\}} + \Ro_{1 \sgoes 2}    & \geq  D(P_{ U_{1 \sgoes 2}^{\rm e} U_{1 \sgoes \{1,2\}}^{\rm e} } ||P_{ U_{1 \sgoes 2}^{\rm c} U_{1 \sgoes \{1,2\}}^{\rm c} } ) \nonumber \\
 \Ro_{1 \sgoes \{1,2\}} \ \quad \quad \quad	             & \geq   D(P_{U_{1 \sgoes \{1,2\}}^{\rm e} } ||P_{U_{1 \sgoes \{1,2\}}^{\rm c} }) \nonumber \\
 L_{1 \sgoes \{1,2\}}+L_{1 \sgoes 1} & \leq  I(Y_1 ; U_{1 \sgoes \{1,2\}}^{\rm e} U_{1 \sgoes 1}^{\rm e})  \nonumber \\
& \quad \quad + D( P_{U_{1 \sgoes 1}^{\rm e}U_{1 \sgoes \{1,2\}}^{\rm e} }|| P_{U_{1 \sgoes 1}^{\rm c}  U_{1 \sgoes \{1,2\}}^{\rm c} })\nonumber \\
 L_{1 \sgoes 1} & \leq  I(Y_1 ;  U_{1 \sgoes 1}^{\rm e}| U_{1 \sgoes \{1,2\}}^{\rm e}) \nonumber \\
& \quad \quad + D( P_{U_{1 \sgoes 1}^{\rm e} }|| P_{U_{1 \sgoes 1}^{\rm c}  } | P_{U_{1 \sgoes \{1,2\}}^{\rm e}}) \nonumber \\
 L_{1 \sgoes \{1,2\}}+L_{1 \sgoes 2}  & \leq  I(Y_2 ; U_{1 \sgoes \{1,2\}}^{\rm e} U_{1 \sgoes 2}^{\rm e}) \nonumber \\
& \quad \quad +D( P_{U_{1 \sgoes 2}^{\rm e} U_{1 \sgoes \{1,2\}}^{\rm e} }|| P_{U_{1 \sgoes 2}^{\rm c} U_{1 \sgoes \{1,2\}}^{\rm c} })\nonumber \\
 L_{1 \sgoes 2} & \leq  I(Y_2 ;  U_{1 \sgoes 2}^{\rm e}| U_{1 \sgoes \{1,2\}}^{\rm e}) \nonumber \\
& \quad \quad + D( P_{U_{1 \sgoes 2} }^{\rm e}|| P_{U_{1 \sgoes 2}^{\rm c}  } | P_{U_{1 \sgoes \{1,2\}}^{\rm e}})
}{\label{eq:An achievable region for the BC-CM}}
union over all the distributions that factor as
\eas{
P_{\rm codebook} &= P_{U_{ 1 \sgoes \{1,2\} }^{\rm c}} P_{U_{1 \sgoes 1 }^{\rm c} | U_{ 1 \sgoes \{1,2\}}^{\rm c}} P_{U_{1 \sgoes 2 }^{\rm c} | U_{ 1 \sgoes \{1,2\}}^{\rm c}} \\
P_{\rm encoding} &= P_{U_{1 \sgoes 1 }^{\rm e} U_{1 \sgoes 2 }^{\rm e} U_{ 1 \sgoes \{1,2\}}^{\rm e}},
}{\label{eq:distribution BC}}
for the rate splitting strategy in \eqref{eq:rate splitting BC} and $L_{\iv \sgoes \jv}=R_{\iv \sgoes \jv}+\Ro_{\iv \sgoes \jv}$.
\end{cor}

\begin{IEEEproof}
 The complete proof is provided in \cite{rini2012Extension}.
\end{IEEEproof}

After the FME of the binning rates
$\Ro_{\iv \sgoes \jv}$
 we obtain the region
\ea{
R_{1 \sgoes 1} \quad \quad  \quad \ & \leq  I(Y_1 ;  U_{1 \sgoes 1}^{\rm e}| U_{1 \sgoes \{1,2\}}^{\rm e})\nonumber \\
&  + D( P_{U_{1 \sgoes 1}^{\rm e} }|| P_{U_{1 \sgoes 1}^{\rm c}  } | P_{U_{1 \sgoes \{1,2\}}}^{\rm e}) \nonumber \\
R_{1 \sgoes 2} & \leq  I(Y_2 ;  U_{1 \sgoes 2}^{\rm e}| U_{1 \sgoes \{1,2\}}^{\rm e}) \nonumber \\
& + D( P_{U_{1 \sgoes 2}^{\rm e} }|| P_{U_{1 \sgoes 2}^{\rm c}  } | P_{U_{1 \sgoes \{1,2\}}^{\rm e}} ) \nonumber \\
R_{1 \sgoes \{1,2\}}+R_{1 \sgoes 1} \quad \quad  \quad \ & \leq  I(Y_1 ; U_{1 \sgoes \{1,2\}}^{\rm e} U_{1 \sgoes 1}^{\rm e})\nonumber \\
R_{1 \sgoes \{1,2\}}\quad \quad  \quad \ +R_{1 \sgoes 2} & \leq  I(Y_2 ; U_{1 \sgoes \{1,2\}}^{\rm e} U_{1 \sgoes 2}^{\rm e})\nonumber \\
R_{1 \sgoes \{1,2\}}+R_{1 \sgoes 1}+ R_{1 \sgoes 2} & \leq  I(Y_1 ; U_{1 \sgoes \{1,2\}}^{\rm e} U_{1 \sgoes 1}^{\rm e})\nonumber \\
\quad \quad +I(Y_2 ; U_{1 \sgoes 2}^{\rm e} | U_{1 \sgoes \{1,2\}}^{\rm e})&-D_{\rm BC-RS}\nonumber \\
R_{1 \sgoes \{1,2\}}+R_{1 \sgoes 1}+ R_{1 \sgoes 2} & \leq  I(Y_2 ; U_{1 \sgoes \{1,2\}}^{\rm e} U_{1 \sgoes 2}^{\rm e})\nonumber \\
\quad \quad +I(Y_1 ; U_{1 \sgoes 1}^{\rm e} | U_{1 \sgoes \{1,2\}}^{\rm e}) &-D_{\rm BC-RS},
\label{eq:An achievable region for the BC-CM}
}
with
\ea{
& D_{\rm BC-RS} =
 \sum_{u_{1 \sgoes 1}, u_{1 \sgoes 2}, u_{1 \sgoes \{1,2\}}} P_{U_{1 \sgoes \{1,2\}}^{\rm e} U_{1 \sgoes 1}^{\rm e} U_{1 \sgoes 2}^{\rm e}} \nonumber \\
& \quad \quad \cdot  \log \lb \f{  P_{U_{1 \sgoes 1}^{\rm e}| U_{1 \sgoes \{1,2\}}^{\rm e}} }
 { P_{U_{1 \sgoes 1}^{\rm e} |U_{1 \sgoes \{1,2\}}^{\rm e} U_{1 \sgoes 2}^{\rm e}} } \f {  P_{U_{1 \sgoes 1}^{\rm c} |U_{1 \sgoes \{1,2\}}^{\rm c} U_{1 \sgoes 2}^{\rm c}} }
{P_{U_{1 \sgoes 1}^{\rm c} | U_{1 \sgoes \{1,2\}}^{\rm c}} }  \rb \nonumber  \\
& \quad \quad \quad  \ \ = - I(U_{1 \sgoes 1}^{\rm e} , U_{1 \sgoes 2}^{\rm e}| U_{1 \sgoes \{1,2\}}^{\rm e}).
\label{eq:DBC-CM}
}
With the equivalence in \eqref{eq:DBC-CM} we conclude that the region in \eqref{eq:An achievable region for the BC-CM} is equivalent to Marton's region
\cite{marton1979coding} which is the largest known achievable region for a general BC.
As for Cor. \ref{cor:An Achievable Region for the MAC}, Cor. \ref{cor:An Achievable Scheme for the BC} shows that the achievable region is not determined
by the distribution of the codewords in the codebook but only on the distribution after encoding.
%
%

\subsection{The Interference Channel}
\label{sec:IFC}

The IFC is four-terminal network where two pairs of transmitter/receiver pairs want to communicate a message over the channel each.
As for the BC, we can rate-split each message into a public and private part: the private messages, $W_{1 \sgoes 1}$ and $W_{2\sgoes 2}$ respectively, are decoded
only at the intended transmitter while the public messages, $W_{1 \sgoes \{1,2\}}$ and $W_{2\sgoes \{1,2\}}$, are decoded by both decoders.
Rate-splitting transforms the problem of achieving the rate vector $(R_{1 \sgoes 1}',R_{2 \sgoes 2}')$ in the problem of achieving
the rate vector $(R_{1 \sgoes 1},R_{2 \sgoes 2}, R_{1 \sgoes \{1,2\}}, R_{2\sgoes \{1,2\}} )$  where
\ea{
\small
\lsb \p {
R_{1 \sgoes 1}' \\
R_{2 \sgoes 2}'
}\rsb
=
\lsb \p{
\al \ 0 \ \alb \ 0 \\
0  \ \be \ 0   \ \overline{\be} \\
}\rsb \small
\lsb R_{1\sgoes 1 }  \ R_{2\sgoes 2} \  R_{1\sgoes \{1,2\}} \ R_{2 \sgoes \{1,2\}}\rsb^T,
\label{eq:rate splitting IFC}
}
for any $(\al,\be) \in [0 \ldots 1]^2 $ , $\alb=1-\al$ , $\overline{\be}=1 - \be$.
The new random coding technique of Sec. \ref{sec:Combining Superposition Coding and Binning}
can be applied to the IFC after rate splitting by superposing the $U_{1 \sgoes 1 }^N$ onto  $U_{1 \sgoes \{1,2\}}^N$ and
jointly binning $U_{1 \sgoes 1}^N,U_{1 \sgoes \{1,2\}}^N$. The same encoding procedure is applied to the codewords $U_{ 2 \sgoes 2 }^N$ and
$U_{ 2 \sgoes \{1,2\}}^N$.

\begin{cor}{\bf An Achievable Scheme for the IFC}
\label{cor:An Achievable Scheme for the IFC}
The following region is achievable for a general IFC:
\pp{
\Ro_{1 \sgoes \{1,2\}}  +\Ro_{1\sgoes 1}    \geq  D(P_{U_{1 \sgoes 1}^{\rm e}   , U_{1 \sgoes \{1,2\}}^{\rm e} } || P_{U_{1 \sgoes 1}^{\rm c}   , U_{1 \sgoes \{1,2\}}^{\rm c} }) \nonumber \\
\quad \quad \quad \quad \quad		    \Ro_{1 \sgoes 1}    \geq D(P_{ U_{1 \sgoes 1}^{\rm e}} ||P_{ U_{1 \sgoes 1}^{\rm c}} | P_{ U_{1 \sgoes \{1,2\}}^{\rm e} }  )\nonumber \\
\Ro_{2 \sgoes \{1,2\}}  +\Ro_{2\sgoes 2}  \geq D(P_{U_{2 \sgoes 2}^{\rm e}   , U_{2 \sgoes \{1,2\}}^{\rm e} } || P_{U_{2 \sgoes 2}^{\rm c}   , U_{2 \sgoes \{1,2\}}^{\rm c} }) \nonumber \\
\quad \quad \quad \quad \quad		    \Ro_{2 \sgoes 2}   \geq  D(P_{ U_{2 \sgoes 2}^{\rm e} } ||P_{ U_{2 \sgoes 2}^{\rm c}} | P_{ U_{2 \sgoes \{1,2\}}^{\rm e} } )\nonumber \\
L_{1 \sgoes \{1,2\}}+L_{1 \sgoes 1}+L_{2 \sgoes \{1,2\} }   \leq  I(Y_1 ; U_{1 \sgoes \{1,2\}}^e,  U_{1 \sgoes 1}^e, U_{2 \sgoes \{1,2\}}^e) \nonumber  \nonumber \\
\quad \quad   + D( P_{U_{1 \goes 1}  U_{1 \goes \{1,2\}}}^e  || P_{U_{1 \goes 1} U_{1 \goes \{1,2\}}}^c) \nonumber \\
\quad \quad \quad \quad \ \ L_{1 \sgoes 1}+L_{2 \sgoes \{1,2\}}   \leq  I(Y_1 ;  U_{1 \sgoes 1}^e, U_{2 \sgoes \{1,2\}}^e| U_{1 \sgoes \{1,2\}}^e )  \nonumber \nonumber \\
\quad \quad  + D(P_{U_{1 \goes 1}^{\rm e}} || P_{U_{1 \goes 1}^{\rm c}} | P_{U_{1 \goes \{1,2\}}^{\rm e}}  ) \nonumber \\
%
L_{1 \sgoes \{1,2\}}+L_{1 \sgoes 1} \quad  \quad  \quad  \quad \quad   \leq  I(Y_1 ; U_{1 \sgoes \{1,2\}}^e,  U_{1 \sgoes 1}^e| U_{2 \sgoes \{1,2\}}^e) \nonumber \nonumber \\
\quad \quad    + D( P_{U_{1 \goes 1}^{\rm e}  U_{1 \goes \{1,2\}}^{\rm e}}  || P_{U_{1 \goes 1}^{\rm c} U_{1 \goes \{1,2\}}^{\rm c}}) \nonumber \\
\quad  \quad  \quad \quad \quad L_{1 \sgoes 1}     \quad  \quad \quad  \ \ \ \ \ \leq  I(Y_1 ;  U_{1 \sgoes 1}^e| U_{1 \sgoes \{1,2\}}^e ,  U_{2 \sgoes \{1,2\}}^e)  \nonumber  \nonumber \\
\quad \quad    + D(P_{U_{1 \goes 1}^{\rm e}} || P_{U_{1 \goes 1}^{\rm c}} | P_{U_{1 \goes \{1,2\}}^{\rm e}}  ) \nonumber \\
L_{1 \sgoes \{1,2\}}+L_{2 \sgoes 2}+L_{2 \sgoes \{1,2\}}    \leq  I(Y_2 ; U_{1 \sgoes \{1,2\}}^e,  U_{2 \sgoes 2}^e, U_{2 \sgoes \{1,2\}}^e) \nonumber  \nonumber \\
\quad \quad    + D( P_{U_{2 \goes 2}^{\rm e}  U_{2 \goes \{1,2\}}^{\rm e}}  || P_{U_{2 \goes 2}^{\rm c} U_{2 \goes \{1,2\}}^{\rm c}}) \nonumber \\
\quad \quad  \quad \quad \ \ \ 		     L_{2 \sgoes 2}+L_{2 \sgoes \{1,2\}}  \leq  I(Y_2 ;  U_{2 \sgoes 2}^e, U_{2 \sgoes \{1,2\}}^e| U_{1 \sgoes \{1,2\}}^e )  \nonumber \nonumber \\
\quad \quad    + D(P_{U_{2 \goes 2}^{\rm e} U_{2 \goes \{1,2\}}^{\rm e}} || P_{U_{2 \goes 2}^{\rm c} U_{2 \goes \{1,2\}}^{\rm c}}  ) \nonumber \\
  }
  \pp
  {
L_{1 \sgoes \{1,2\}}+L_{2 \sgoes 2} \quad  \quad  \quad  \leq  I(Y_2 ; U_{1 \sgoes \{1,2\}}^e,  U_{2 \sgoes 2}^e| U_{2 \sgoes \{1,2\}}^e) \nonumber \nonumber \\
\quad \quad    + D( P_{U_{2 \goes 2}^{\rm e}} || P_{U_{2 \goes 2}^{\rm c}} | P_{U_{2 \goes \{1,2\}}^{\rm e}}) \nonumber \\
\quad  \quad  \quad  \quad  \quad   L_{2 \sgoes 2} \quad \quad \quad \leq  I(Y_2 ;  U_{2 \sgoes 2}^e| U_{1 \sgoes \{1,2\}}^e ,  U_{2 \sgoes \{1,2\}}^e)  \nonumber \nonumber \\
}
\vspace{-.65 cm}
\ea{
\hspace{-2.7 cm }
  + D(P_{U_{2 \goes 2}^{\rm e}} || P_{U_{2 \goes 2}^{\rm c}} | P_{U_{2 \goes \{1,2\}}^{\rm e}}  )
}
union over all the distributions that factor as
\ea{
P_{\rm codebook} &= P_{U_{ 1 \sgoes \{1,2\} }^{\rm c}} P_{U_{1 \sgoes 1 }^{\rm c} | U_{ 1 \sgoes \{1,2\}}^{\rm c}} P_{U_{1 \sgoes 2 }^{\rm c} | U_{ 1 \sgoes \{1,2\}}^{\rm c}} \nonumber \\
P_{\rm encoding} &= P_{U_{1 \sgoes 1 }^{\rm e} U_{1 \sgoes 2 }^{\rm e} U_{ 1 \sgoes \{1,2\}}^{\rm e}},
\label{eq:distribution IFC}}
for the rate splitting strategy in \eqref{eq:rate splitting IFC} and $L_{\iv \sgoes \jv}=R_{\iv \sgoes \jv}+\Ro_{\iv \sgoes \jv}$.
\end{cor}

\begin{IEEEproof}
 The complete proof is provided in \cite{rini2012Extension}.
\end{IEEEproof}

From the FME of the binning rates
$\Ro_{\iv \sgoes \jv}$ one obtains that the largest achievable
region  in $(R_{1 \sgoes  1}',R_{2 \sgoes  2}')$ is achieved with the choice $P_{U_{1 \sgoes \{1,2\}}^{\rm e} }=P_{U_{1 \sgoes \{1,2\}}^{\rm c}}$ and
$P_{U_{2 \sgoes \{1,2\}}^{\rm e}}=P_{U_{2 \sgoes \{1,2\}}^{\rm c}}$. With this choice the achievable region in  Cor. \ref{cor:An Achievable Scheme for the IFC}
becomes equivalent to the Han and Kobayashi region \cite{Han_Kobayashi81}, which is the largest known achievable region for a general IFC.

As for the MAC , Cor. \ref{cor:An Achievable Region for the MAC}, and the BC,  Cor. \ref{cor:An Achievable Scheme for the BC},
 Cor. \ref{cor:An Achievable Scheme for the IFC} does not improve on the largest known region for the IFC but shows that a larger set of transmission
 strategies than superposition coding and binning can be used to achieve this region.

\medskip

In \cite[Sec. VI]{rini2011achievable} we have shown that superposition coding and binning can both be used to achieve the largest known inner bound for the MAC, BC
and IFC.
In the examples above we have shown that combining the two encoding strategies into a new and more general transmission strategy still achieves
the same performance.
With these considerations in mind,  we can provide an insight on the error performance of these two coding techniques.
%
%
%
%
%
Both in  superposition coding and binning one creates multiple codewords to transmit the same message.
In superposition coding, the message encoded in top codebook is associated to multiple codewords, one for each possible base codeword.
while, in binning, codewords in the same bin are associated to the same message.
%
%
%
%
When $U_2^N$, with rate $R_2$, is superposed to $U_1^N$, with rate $R_1$, the number of possible codewords used to encode the same message in $U_2^N$ is
$2^{N R_1}$.
In binning, the number of excess codewords depends on the joint probability distribution between the codewords imposed by the encoding procedure.
When $U_2$ is binned against $U_1$, the smallest number of codeword in each bin is $\Ro_2=I(U_1; U_2)$.
%
%
In both cases, the transmitted codewords $[U_1, U_2]$ belong to the typical set $\Tcal_{\ep}^N (P_{U_1 U_2})$ but
superposition coding usually requires a larger number of excess codewords than binning to achieve the desired typicality property
and this number is fixed and does not depend on $P_{U_1 U_2}$.
While binning is more advantageous than superposition coding at encoding, it performs worst at decoding.
%
In superposition coding, after the decoding of the base codeword $U_1^N$, the receiver looks for the transmitted top codeword $U_2^N$ in a codebook
of size $2^{N R_2}$.
In binning, instead, after $U_1^N$ has been correctly decoded, the possible transmitted codewords are $2^{N (R_2 +\Ro_2)}$.
The knowledge of $U_1^N$ helps the decoder in determining $U_2^N$ in that $U_1^N,U_2^N$ must appear as if generated according to the encoding distributions,
but it does not reduce the number of possible transmitted codewords $U_2^N$.
Interestingly the encoding and decoding benefits  provided by superposition coding and binning seem to balance each other in the proposed random coding technique.

\section{Conclusion}
\label{sec:conclusion}

In this paper we present a new achievable strategies that encompasses superposition coding and binning.
The error analysis of this new achievable scheme requires a more general version of the classical covering lemma that is based on the \emph{inaccuracy} between
typical sequences.
With this new random coding technique we derive achievable regions for the multi-access channel, broadcast channel and interference channel.
These inner bounds do not improve on the largest known achievable regions but show that the same error performance can be achieved with
a large set of encoding strategies.

\section*{Acknowledgment}

The author would like to thank Prof. Gerhard Kramer for suggesting the \emph{inaccuracy} to measure the distance between codebook and encoding distribution.

\bibliographystyle{IEEEtran}
\bibliography{steBib1}

\end{document}